\documentstyle[12pt,aps]{revtex}

\widetext
\draft
\tighten

\begin{document}

\preprint{EFUAZ FT-98-63}

\title{{\rm {\bf EL FORMALISMO DE BARGMANN Y WIGNER MODIFICADO:  LOS
ESPINES ALTOS}}\thanks{Enviado a III Reuni\'on Nacional Acad\'emica de
F\'{\i}sica y Matem\'aticas, ESFM IPN, D. F., M\'exico, Agosto 3-7, 1998 y
presentado en la XXXI Escuela Latinoamericana de F\'{\i}sica, El Colegio
Nacional, D.  F., M\'exico, Julio 27 -- Agosto 14, 1998.}}

\author{{\bf Valeri V. Dvoeglazov}}

\address{{\rm
Escuela de F\'{\i}sica, Universidad Aut\'onoma de Zacatecas \\
Apartado Postal C-580, Zacatecas 98068, ZAC., M\'exico\\
Email:  valeri@cantera.reduaz.mx\\
URL: http://cantera.reduaz.mx/\~\,valeri/valeri.htm}
}


\maketitle

\bigskip

\begin{abstract}
En los art\'{\i}culos anteriores de Ogievetski\u{\i} y Polubarinov,
Kalb y Ramond el concepto del {\it not\'of}, el campo longitudinal
originado del tensor antisim\'etrico (AST), ha sido
propuesto.  En nuestro trabajo analisamos la teor\'{\i}a del campo AST
de segundo rango desde el punto de vista del problema de la
normalizaci\'on.  Obtenemos 4-potenciales y fuerzas de campo, que
coinciden con los que fueron  obtenidos en los trabajos de Ahluwalia y
Dvoeglazov.  Ligeramente modificando la funci\'on de campo de
Bargmann-Wigner (BW) concluimos que es posible  describir
expl\'{\i}citamente
grados de libertad del fot\'on y el {\it not\'of}.  Se discuten las
consecuencias f\'{\i}sicas y relaciones a los trabajos antecedentes.
Adem\'as, derivamos ecuaciones para el tensor {\it sim\'etrico} de
segundo rango en la base de la misma modificaci\'on del formalismo de
Bargmann y Wigner, i.e.  las ecuaciones que describen conducta din\'amica
de los campos con spin m\'aximo 2.
\end{abstract}

\pacs{PACS: 03.65.Pm, 04.50.+h, 11.30.Cp}


El esquema general para derivaci\'on de ecuaciones de los espines
altos ha sido dada en~\cite{BW}.  Un campo de masa $m$ y spin
$j \geq {1\over  2}$ es representado por un multispinor
completamente sim\'etrico del rango $2j$. Los casos
particulares $j= 1$ y $j= {3\over 2}$ han sido dados en los libros
de texto, e.~g., ref.~\cite{Lurie}. En el presente trabajo nosotros
aprobamos que el esquema podr\'{\i}a ser generalizada para el caso de spin
1 y $3/2$.  El  caso spin-2 puede ser tambi\'en de unos inter\'es, porque
generalmente se cree que se obtienen de componentes transversos los rasgos
esenciales del campo gravitacional de la representaci\'on $ (2,0)\oplus
(0,2) $ del grupo de Lorentz. No obstante, todav\'{\i}a no se entienden
preguntas de las componentes redundantes  de las ecuaciones relativistas
de los espines altos en detalles~\cite{Kirch}.

En una serie reciente de los papeles~\cite{DVO0,DVO1,DVO2},
que son continuaci\'on de los trabajos de Ahluwalia (vease, por
ejemplo,~\cite{DVA0,DVA1,DVA2} tratamos de construir
una teor\'{\i}a consistente del tensor antisim\'etrico del campo
cuantizado de segundo rango y del campo 4-vectorial. Los trabajos
publicados previamente~\cite{OP,HA,KR}, as\'{\i}
como libros de texto, no pueden considerarse
como los trabajos que resolvieron los problemas principales,
si el campo AST  cuantizado  y el campo 4-vectorial cuantizado
son los campos ``transversos" \'o ``longitudinales" (en el sentido si la
helicidad $h=\pm 1$ \'o $h= 0$? pueden  los potenciales
electromagn\'eticos componer un 4-vector en una teor\'{\i}a cuantizada?
c\'omo se debe tomar el l\'{\i}mite sin masa? etc. etc.
Muchos problemas de la descripci\'on rigurosa de la luz son
todav\'{\i}a abiertos. Se basan ideas de este art\'{\i}culo en tres
informes del \'arbitro de`` Foundation of Physics" que eran muy \'utiles
aunque cr\'{\i}ticos.

Primeros de todo, notamos despu\'es del \'arbitro que 1)``...En unidades
naturales ($c=\hbar=1$) una densidad de lagrangiano (porque la acci\'on
se considera  sin dimensi\'on), tiene dimensi\'on de [energ\'{\i}a]$^4$";
2) Uno puede siempre re-normalizar la densidad del lagrangiano y ``uno
puede obtener las mismas ecuaciones de movimiento ... por sustituir $L\
\rightarrow (1/M^N) L$, donde $M$ es una escala de energ\'{\i}a arbitraria,"
cf.~\cite{DVO1}; 3) la dimensi\'on f\'{\i}sica correcta del
tensor de la intensidad del campo $F^{\mu\nu}$ es [energ\'{\i}a]$^2$;
``la transformaci\'on $F^{\mu\nu}\rightarrow (1/2m) F^{\mu\nu}$
(v\'ease en ref.~[5,6a]) requiere un estudio m\'as detallado,
[porque] la transformaci\'on mencionada cambia su dimensi\'on
f\'{\i}sica:  no es una transformaci\'on de la normalizaci\'on simple."
Adem\'as, en los primeros art\'{\i}culos del {\it not\'of}~\cite{OP,HA,KR}
\footnote{Tambi\'en  se conoce como un campo longitudinal de Kalb y Ramond,
pero la consideraci\'on de Ogievetski\u{\i} y Polubarinov me
parece ser m\'as rigurosa porque permite estudiar el limite
$m\rightarrow 0$.} los autores usaron la normalizaci\'on del
campo vectorial $F^\mu$ \footnote{Se sabe bien que
est\'a relacionado con un tensor del campo antisim\'etrico de
tercer rango.} a [energ\'{\i}a]$^2$ y, por lo tanto, los potenciales
tensoriales $A^{\mu\nu}$, a [energ\'{\i}a]$^1$.

Despu\'es de tener en cuenta estas observaciones perm\'{\i}tanos
repetir el procedimiento de las derivaciones de las ecuaciones de
Proca desde las ecuaciones de  Bargmann y Wigner  para un spinor
totalmente {\it sim\'etrico} del segundo rango. Ponemos
\begin{equation}
\Psi_{\{\alpha\beta\}} = (\gamma^\mu R)_{\alpha\beta} (c_a m A_\mu +
c_f  F_\mu) +(\sigma^{\mu\nu} R)_{\alpha\beta} (c_A m\gamma^5
A_{\mu\nu} + c_F F_{\mu\nu})\, ,\label{si}
\end{equation}
donde
\begin{equation}
R=\pmatrix{i\Theta & 0\cr 0&-i\Theta\cr}\quad,\quad \Theta = -i\sigma_2 =
\pmatrix{0&-1\cr 1&0\cr}\, .
\end{equation}

Matrices $\gamma^{\mu}$ se escogen en la representaci\'on de  Weyl,
i.e., $\gamma^5$ se asume que sea diagonal.
Las constantes $c_i$  son algunos coeficientes num\'ericos sin dimensi\'on.
El operador de reflexi\'on $R$ tiene las propiedades siguientes
\begin{mathletters}
\begin{eqnarray}
&& R^T = -R\,,\quad R^\dagger =R = R^{-1}\,,\quad
R^{-1} \gamma^5 R = (\gamma^5)^T\,,\\
&& R^{-1}\gamma^\mu R = -(\gamma^\mu)^T\,,\quad
R^{-1} \sigma^{\mu\nu} R = - (\sigma^{\mu\nu})^T\,.
\end{eqnarray}
\end{mathletters}
que son necesarias para la expansi\'on (\ref{si})
sea posible en tal forma, i.e., se suponen que $\gamma^{\mu} R$,
$\sigma^{\mu\nu} R$ y $\gamma^5\sigma^{\mu\nu} R$
son las matrices {\it sim\'etricas}.

La substituci\'on de la expansi\'on precedente en el
conjunto de Bargmann y Wigner~\cite{Lurie}
\begin{mathletters}
\begin{eqnarray}
\left [ i\gamma^\mu
\partial_\mu -m \right ]_{\alpha\beta} \Psi_{\{\beta\gamma\}} (x) &=&
0\,,\label{bw1}\\
\left [ i\gamma^\mu \partial_\mu -m \right
]_{\gamma\beta} \Psi_{\{\alpha\beta\}} (x) &=& 0\, . \label{bw2}
\end{eqnarray}
\end{mathletters}
nos da las
nuevas ecuaciones de Proca:
\begin{mathletters}
\begin{eqnarray}
&& c_a m (\partial_\mu A_\nu - \partial_\nu A_\mu ) +
c_f (\partial_\mu F_\nu -\partial_\nu F_\mu ) =
ic_A m^2 \epsilon_{\alpha\beta\mu\nu} A^{\alpha\beta} +
2 m c_F F_{\mu\nu} \, \label{pr1} \\
&& c_a m^2 A_\mu + c_f m F_\mu =
i c_A m \epsilon_{\mu\nu\alpha\beta} \partial^\nu A^{\alpha\beta} +
2 c_F \partial^\nu F_{\mu\nu}\, . \label{pr2}
\end{eqnarray}
\end{mathletters}
En el caso $c_a= 1$, $c_F={ 1\over 2} $ y $c_f= c_A= 0$ se
reducen a las ecuaciones de Proca ordinarias.\footnote{Sin embargo, notamos
que la divisi\'on por $m$ en la primera ecuaci\'on {\it no} es una
operaci\'on bien-definida en el caso de que  alguien se interese por
el l\'{\i}mite $m\rightarrow 0$ posteriormente.  Probablemente,
para evitar este punto oscuro, uno desear\'{\i}a escribir la ecuaci\'on
de Dirac en la forma $\left [ (i\gamma^\mu
\partial_\mu)/m - \openone \right ] \psi (x) =0$
la que sigue imediatamente en la
derivaci\'on de la ecuaci\'on de Dirac  en la base a la relaci\'on de
Ryder y Burgard~\cite{DVA0} y las reglas de  Wigner para un {\it boost}
de la funci\'on del campo desde el sistema con momento lineal cero.}
En el caso generalizado obtenemos ecuaciones din\'amicas que
conectan el fot\'on, el {\it not\'of} y sus potenciales.
Partes divergentes (en  $m\rightarrow 0$) de funciones del campo
y de variables din\'amicas se deben quitar por transformaciones de
calibraci\'on correspondientes (o de transformaciones de
calibraci\'on de Kalb y Ramond).
Se sabe bien que el campo del {\it not\'of} sin masa resulta ser
puro campo longitudinal cuando se
tiene en cuenta $\partial_\mu A^{\mu\nu}= 0$.

Aparte de estas ecuaciones din\'amicas, podemos obtener el juego de
restricciones por medio de la substracci\'on de las ecuaciones de
Bargmann y Wigner (en lugar de la suma en cuanto
a (\ref{pr1},\ref{pr2})). Se leen
\begin{equation}
mc_a \partial^\mu A_\mu + c_f \partial^\mu f_\mu =0\, ,\mbox{y}
\quad
mc_A \partial^\alpha A_{\alpha\mu} + {i\over 2}
c_F \epsilon_{\alpha\beta\nu\mu}
\partial^\alpha F^{\beta\nu} = 0\, .
\end{equation}
que sugieren
$\widetilde F^{\mu\nu}\sim im A^{\mu\nu}$ y $f^\mu\sim m A^\mu$, como
en~\cite{OP}.

Nosotros  volvimos a estos trabajos antiguos
debido a controversias recientes de interpretaci\'on despues
de observaciones experimentales de los objetos ${\bf E}\times
{\bf E}^\ast$ y ${\bf A}\times {\bf A}^\ast$ en  \'optica
no-lineal. En esta
conexi\'on uno puede considerar que $\sim {\bf A}\times{\bf
A}^\ast$ se puede mirar como la parte del {\it potencial} tensorial
y $\sim {\bf B}\times {\bf B}^\ast$, como la parte del campo
vectorial (cf.  las f\'ormulas (19a-c) en
ref.~[6a]).  Seg\'un~\cite[Eqs. (9,10)]{OP} procedemos en la
construcci\'on de los ``potenciales" para el {\it not\'of} como sigue:
$A_{\mu\nu} ({\bf p})  \sim \left
[\epsilon_\mu^{(1)} ({\bf p})\epsilon_\nu^{(2)} ({\bf p})-
\epsilon_\nu^{(1)} ({\bf p}) \epsilon_\mu^{(2)} ({\bf p}) \right ]$
Al usar formas expl\'{\i}citas de los vectores de polarizaci\'on en el
espacio de momento lineal (e.~g., refs.~\cite{Wein} y~[6a,f\'ormulas
(15a, b)]) uno obtiene
\begin{eqnarray} A^{\mu\nu} ({\bf p}) = {iN^2 \over m} \pmatrix{0&-p_2&
p_1& 0\cr p_2 &0& m+{p_r p_l\over p_0+m} & {p_2 p_3\over p_0 +m}\cr -p_1 &
-m - {p_r p_l \over p_0+m}& 0& -{p_1 p_3\over p_0 +m}\cr 0& -{p_2 p_3
\over p_0 +m} & {p_1 p_3 \over p_0+m}&0\cr}\, , \label{lc}
\end{eqnarray}
que coincide con las componentes
``longitudinales" del tensor antisim\'etrico que fue obtenido en
refs.~\cite[Eqs. (2.14,2.17)]{DVA1} y~[6a,Eqs.(17b,18b)]
dentro de la normalizaci\'on y formas diferentes de la base de spin.
Los estados longitudinales pueden reducirse a cero en el caso sin masa
bajo opci\'on apropiada de la normalizaci\'on y si una
part\'{\i}cula de spin $j=1$ se mueve en direcci\'on  del tercer eje $OZ$.
Tambi\'en es util comparar Eq. (\ref{lc}) con la f\'ormula (B2) en
ref.~\cite{DVA2} para dar cuenta del procedimiento correcto para toma
el l\'{\i}mite sin masa.

Como discusi\'on queremos mencionar que el experimento de Tam y Happer
~\cite{TH} no hac\'{\i}a hallazgo una explicaci\'on satisfactoria en el
armaz\'on de la electrodin\'amica cu\'antica ordinaria (por lo menos, su
explicaci\'on es complicada por grandes c\'alculos  t\'ecnicos). En
cambio, en ref.~\cite{Prad} se ha propuesto un modelo muy
interesante.  Se basa en calibrar el  campo de Dirac usando los
par\'ametros dependientes de coordenadas $\alpha_{\mu\nu} (x) $
en
$\psi(x) \rightarrow \psi^\prime (x^\prime) = \Omega \psi(x)\,\,, \quad
\Omega = \exp \left [ {i\over 2} \sigma^{\mu\nu} \alpha_{\mu\nu}(x)
\right ]\, $
y, as\'{\i}, el segundo ``fot\'on" fue introducido.
El campo de compensaci\'on con  24 componentes (en general)
$B_{\mu,\nu\lambda}$ reduce al campo 4-vectorial como sigue (la
anotaci\'on de~\cite{Prad} se usa aqu\'{\i}):
$B_{\mu,\nu\lambda} = {1\over 4} \epsilon_{\mu\nu\lambda\sigma} a_\sigma
(x) \, .$
Como  prontamente ve despu\'es de la comparaci\'on de estas f\'ormulas con
las de refs.~\cite{OP,HA,KR}, el segundo fot\'on es nada m\'as que
el {\it not\'of} de Ogievetski\u{\i} y Polubarinov dentro de
la normalizaci\'on.  Propiedades de la paridad (as\'{\i} como su conducta
en el l\'{\i}mite sin masa) son dependientes no s\'olo de las formas
expl\'{\i}citas de las funciones del campo en el espacio del momento lineal
de la representaci\'on $(1/2,1/2)$ , {\it sino} tambi\'en
de las propiedades de las operadores de creaci\'on y
aniquilaci\'on correspondientes.  Las propiedades de helicidad
dependen de la normalizaci\'on.

Podemos generalizar el formalismo de otra manera.
En los trabajos~\cite{ras} las generalizaciones de las ecuaciones de
Dirac fueron propuestas. Una de ellas tiene la forma
\begin{equation}
\left [
i\gamma^\mu \partial_\mu - m_1 - \gamma^5 m_2 \right ] \Psi (x) =0\,
\label{rs}
\end{equation}
con dos parametros de masa y permiten describir los estados taquionicos.
Si uno quisiera construir el formalismo para espines altos en base
a la ecuacion (\ref{rs}) llegamos al siguiente sistema de
ecuaciones de Proca para spin 1:\footnote{Para espin 3/2 las
ecuaciones son muy parecidas. Es necesario \'unicamente tomar en cuenta que
la funci\'on con espin 3/2 lleva un \'{\i}ndice adicional relacionado con
la parte espinoreal de ella. Sin embargo, aparecen nuevas restricciones
gracias a un procedimiento de simetrizaci\'on del espinor de tercer rango,
v\'ease~\cite{Lurie}.}
\begin{mathletters}
\begin{eqnarray}
&& 2c_1 \partial_\mu \widetilde F^{\mu\alpha} -2ic_2 \partial_\mu
A^{\mu\alpha}  + m_2 \Psi^\alpha =0\,,\\
&& 2c_1 \partial_\mu F^{\mu\alpha} +2ic_2  \partial_\mu
\widetilde A^{\mu\alpha} + m_1 \Psi^\alpha =0\,,\\
&& 2c_1 (m_1 F^{\mu\nu} + im_2 \widetilde F^{\mu\nu}) +2c_2 (m_2
A^{\mu\nu} +i m_1 \widetilde A^{\mu\nu}) - (\partial^\mu \Psi^\nu -
\partial^\nu \Psi^\mu ) = 0\, ,\\
&& \partial_\mu \Psi^\mu =0\, .
\end{eqnarray} \end{mathletters}
La funci\'on de campo esta representada
por
\begin{equation} \Psi_{\{\alpha\beta\}} = (\gamma^\mu R)_{\alpha\beta}
\Psi_\mu + c_1 (\sigma^{\mu\nu} R)_{\alpha\beta} F_{\mu\nu} +c_2 (\gamma^5
\sigma^{\mu\nu} R)_{\alpha\beta} A_{\mu\nu}\,.  \end{equation} El uso de
dos parametros de masa puede ser \'util para la consideraci\'on del vector
de Pauli y Lubanski (v\'ease~[6a,Eq. (27,28)])

Al repetir la formulaci\'on generalizada para un spinor sim\'etrico de
cuarto rango podemos obtener las siguientes ecuaciones din\'amicas
\begin{mathletters} \begin{eqnarray}
&& {2\alpha_2
\beta_4 \over m} \partial_\nu T_\kappa^{\quad\mu\nu} +{i\alpha_3
\beta_7 \over m} \epsilon^{\mu\nu\alpha\beta} \partial_\nu
\widetilde T_{\kappa,\alpha\beta} = \alpha_1 \beta_1
G_\kappa^{\quad\mu}\,, \label{b}\\
&&{2\alpha_2 \beta_5 \over m} \partial_\nu
R_{\kappa\tau}^{\quad\mu\nu} +{i\alpha_2 \beta_6 \over m}
\epsilon_{\alpha\beta\kappa\tau} \partial_\nu \widetilde R^{\alpha\beta,
\mu\nu} +{i\alpha_3 \beta_8 \over m}
\epsilon^{\mu\nu\alpha\beta}\partial_\nu \widetilde
D_{\kappa\tau,\alpha\beta} - \nonumber\\
&-&{\alpha_3 \beta_9 \over 2}
\epsilon^{\mu\nu\alpha\beta} \epsilon_{\lambda\delta\kappa\tau}
D^{\lambda\delta}_{\quad \alpha\beta} = \alpha_1 \beta_2
F_{\kappa\tau}^{\quad\mu} + {i\alpha_1 \beta_3 \over 2}
\epsilon_{\alpha\beta\kappa\tau} \widetilde F^{\alpha\beta,\mu}\,, \\
&& 2\alpha_2 \beta_4 T_\kappa^{\quad\mu\nu} +i\alpha_3 \beta_7
\epsilon^{\alpha\beta\mu\nu} \widetilde T_{\kappa,\alpha\beta}
=  {\alpha_1 \beta_1 \over m} (\partial^\mu G_\kappa^{\quad \nu}
- \partial^\nu G_\kappa^{\quad\mu})\,, \\
&& 2\alpha_2 \beta_5 R_{\kappa\tau}^{\quad\mu\nu} +i\alpha_3 \beta_8
\epsilon^{\alpha\beta\mu\nu} \widetilde D_{\kappa\tau,\alpha\beta}
+i\alpha_2 \beta_6 \epsilon_{\alpha\beta\kappa\tau} \widetilde
R^{\alpha\beta,\mu\nu}
- {\alpha_3 \beta_9\over 2} \epsilon^{\alpha\beta\mu\nu}
\epsilon_{\lambda\delta\kappa\tau} D^{\lambda\delta}_{\quad \alpha\beta}
= \nonumber\\
&=& {\alpha_1 \beta_2 \over m} (\partial^\mu F_{\kappa\tau}^{\quad \nu}
-\partial^\nu F_{\kappa\tau}^{\quad\mu} ) + {i\alpha_1 \beta_3 \over 2m}
\epsilon_{\alpha\beta\kappa\tau} (\partial^\mu \widetilde
F^{\alpha\beta,\nu} - \partial^\nu \widetilde F^{\alpha\beta,\mu} )\, .
\label{f}
\end{eqnarray}
\end{mathletters}
Las restricciones  esenciales fueron obtenidos en~[6c].
\'Estas resultan de las contracciones de la funci\'on de campo
con tres matrices antisim\'etricas, como anteriormente se hizo en el
formalismo estandar (vease~\cite{Lurie}, el caso de espin $3/2$).

Adem\'as, del juego de ecuaciones (\ref{b}-\ref{f}) uno
obtiene una ecuaci\'on de {\it segundo}  orden para tensor
sim\'etrico de segundo rango ($\alpha_1\neq 0$, $\beta_1\neq 0$):
\begin{equation}
{1\over m^2} \left [\partial_\nu
\partial^\mu G_\kappa^{\quad \nu} - \partial_\nu \partial^\nu
G_\kappa^{\quad\mu} \right ] =  G_\kappa^{\quad \mu}\, .
\end{equation}
Despu\'es de contracci\'on en los indices $\kappa$ y $\mu$ esta ecuaci\'on
se reduce al conjunto
\begin{eqnarray}
\cases{\partial_\mu G_{\quad\kappa}^{\mu} = F_\kappa\,  &\cr
{1\over m^2} \partial_\kappa F^\kappa = 0&}\, ,
\end{eqnarray}
i.~e.,  a las ecuaciones que conectan el an\'alogo del tensor de
energ\'{\i}a-momento y el an\'alogo del potencial cuatridimensional.

Como conclusi\'on, podemos decir que 1) aprobemos la significancia
f\'{\i}sica de la normalizaci\'on; 2) aprobemos que conceptos del {\it
not\'of}, del campo longitudinal de Kalb y Ramond, del torsi\'on etc
est\'an conectados entre s\'{\i} y est\'an relacionados con el componente
de helicidad 0 del campo antisim\'etrico de segundo rango; 3) aprobemos
que se pueden describir los grados de libertad ``transversos" y
``longitudinales" por la misma ecuaci\'on con coefficientes numericos
desconocidos~[6b]; 4) los resultados tienen que ser aplicados a casos de
espin 2 (y, presumidamente, a los casos con espines m\'as altos) por
raz\'on que el formalismo estandar tiene problemas con interpretaci\'on de
las soluciones~[6c].

\bigskip

{\it Agradecimientos.} Les agradezco mucho las  communicaciones por
Internet de las colegas en el mundo.  Particularmente, quisiera mencionar
las discusiones con doctores D.  V. Ahluwalia, A. E. Chubykalo, M.
Israelit, M. Kirchbach y A.  S.  Shumovsky~\cite{Shumovsky}.   Le
agradezco ayuda del Sr. Jes\'us C\'azares Montes en la gramatica
espa\~nola.  El trabajo ha sido apoyado parcialmente por el Sistema
Nacional de Investigadores de M\'exico y Proyecto de CONACyT 0270P-E.

\bigskip

{\bf Apendice.} Se deducen las funciones del campo de la representaci\'on
$(1/2,1/2)$ en el espacio del momento lineal (vease ref.~[6a]):
\begin{mathletters} \begin{eqnarray}
u^\mu ({\bf p}, +1)= -{N\over
m\sqrt{2}}\pmatrix{p_r\cr m+ {p_1 p_r \over p_0+m}\cr im +{p_2 p_r \over
p_0+m}\cr {p_3 p_r \over p_0+m}\cr}&\quad&,\quad u^\mu ({\bf p}, -1)=
{N\over m \sqrt{2}}\pmatrix{p_l\cr m+ {p_1 p_l \over p_0+m}\cr -im +{p_2
p_l \over p_0+m}\cr {p_3 p_l \over p_0+m}\cr}\quad,\quad\\ u^\mu ({\bf p},
0) = {N\over m} \pmatrix{p_3\cr {p_1 p_3 \over p_0+m}\cr {p_2 p_3 \over
p_0+m}\cr m + {p_3^2 \over p_0+m}\cr}&\quad&, \quad u^\mu ({\bf p}, 0_t) =
{N \over m} \pmatrix{p_0\cr p_1 \cr p_2\cr p_3\cr}\quad.  \end{eqnarray}
\end{mathletters}
Estas no son divergentes en el l\'{\i}mite sin masa
si tomamos $N=m$. Las describen los ``fotones" longitudinales (en el
sentido que $h=0$) en este l\'{\i}mite.  Si $N=1$ tenemos
``fotones transversos" pero tenemos tambi\'en la conducta divergente por
parte de la calibraci\'on de las funciones del campo. Se obtienen los
potenciales para las soluciones de energ\'{\i}a negativa por aplicaci\'on
del operador de conjugaci\'on compleja.

Las fuerzas del campo (strengths) son
\begin{mathletters}
\begin{eqnarray}
{\bf B}^{(+)} ({\bf p}, +1) &=& -{iN\over 2\sqrt{2} m} \pmatrix{-ip_3 \cr
p_3 \cr ip_r\cr}\,,\quad
{\bf E}^{(+)} ({\bf p}, +1) =  -{iN\over 2\sqrt{2} m} \pmatrix{p_0- {p_1
p_r \over p_0+m}\cr ip_0 -{p_2 p_r \over p_0+m}\cr -{p_3 p_r \over
p_0+m}\cr} \,,\\
{\bf B}^{(+)} ({\bf p}, 0) &=& {iN \over 2m}
\pmatrix{p_2 \cr -p_1 \cr 0\cr}\,,\quad
{\bf E}^{(+)} ({\bf p}, 0) =  {iN \over 2m} \pmatrix{- {p_1 p_3
\over p_0+m}\cr -{p_2 p_3 \over p_0+m}\cr p_0-{p_3^2 \over
p_0+m}\cr} \,,\\
{\bf B}^{(+)} ({\bf p},
-1) &=& {iN \over 2\sqrt{2}m} \pmatrix{ip_3 \cr p_3 \cr
-ip_l\cr} \,,\quad
{\bf E}^{(+)} ({\bf p}, -1) =  {iN\over 2\sqrt{2} m} \pmatrix{p_0- {p_1
p_l \over p_0+m}\cr -ip_0 -{p_2 p_l \over p_0+m}\cr -{p_3 p_l \over
p_0+m}\cr} \,.
\end{eqnarray} \end{mathletters}
Se obtienen por la aplicaci\'on de las formulas:
${\bf B}^{(\pm)} ({\bf p}, h ) =\pm {i\over 2m} {\bf p} \times {\bf
u}^{(\pm )} ({\bf p}, h)$  y ${\bf E}^{(\pm)} ({\bf p}, h) = \pm {i\over
2m} p_0 {\bf u}^{(\pm )} ({\bf p}, h) \mp {i\over 2m} {\bf p} u^{0\,
(\pm )} ({\bf p}, h)$. Es \'util  comparar estas formulas con aquellas
que fueron presentados en~\cite[p.408]{DVA1}  usando una base
de spin diferente.  Se obtienen tambi\'en los productos cruz $\times$
en ref.~[6a] y como vemos son relacionados con la parte de calibraci\'on
($\sim p^\mu$) de las potenciales cuatridimensionales en el
espacio de momento lineal.


\vspace*{-5mm}
\begin{references}
\footnotesize{
\baselineskip13pt
\vspace*{-20mm}

\bibitem{BW} V. Bargmann y E. P. Wigner, Proc. Natl. Acad. Sci. (USA) {\bf
34} (1948) 211.

\bibitem{Lurie} D. Luri\'e, {\it Particles and Fields.} (Interscience
Publisher, New York, 1968), Cap\'{\i}tulo 1.

\bibitem{Kirch} M. Kirchbach, Mod. Phys. Lett. A{\bf 12} (1997) 2373;
ibid. {\bf 12} (1997) 3177; ibid. {\bf 13} (1998) 823.

\bibitem{DVO0} V. V. Dvoeglazov, Helv. Phys. Acta {\bf 70} (1997) 677;
ibid.  686; ibid. 697.

\bibitem{DVO1} V. V. Dvoeglazov, Int. J. Theor. Phys., en impresi\'on
(1998)

\bibitem{DVO2}V. V. Dvoeglazov, Preprint hep-th/9712036, Nov. 1997;
Preprint physics/9804010, Abr. 1998; Preprint math-ph/9805017, Abr.
1998.

\bibitem{DVA0} D. V. Ahluwalia {\it et al.}, Phys. Lett. B{\bf 316}
(1993) 102.

\bibitem{DVA1} D. V. Ahluwalia y D. J. Ernst, Int. J. Mod. Phys. E{\bf
2} (1993) 5161.

\bibitem{DVA2} D. V. Ahluwalia y M. Sawicki, Phys. Rev. D{\bf 47} (1993)
5161.

\bibitem{OP}  V. I. Ogievetski\u{\i} y I. V. Polubarinov, Sov. J. Nucl.
Phys.  {\bf 4} (1967) 156.

\bibitem{HA} K. Hayashi, Phys. Lett. B{\bf 44} (1973) 497.

\bibitem{KR} M. Kalb y P. Ramond, Phys. Rev. D{\bf 9} (1974) 2273.

\bibitem{Wein}S. Weinberg, {\it Quantum Field Theory. Vol. I. Foundations.}
(Cambridge University Press, 1995), Cap\'{\i}tulo 5.

\bibitem{TH} A. C. Tam y W. Happer, Phys. Rev. Lett. {\bf 38} (1977) 278.

\bibitem{Prad} P. C. Naik y T. Pradhan, J. Phys. A{\bf 14} (1981) 2795;
T. Pradhan {\it et al.}, Pramana J. Phys. {\bf 24} (1985) 77.

\bibitem{ras} M. Markov, ZhETF {\bf 7} (1937) 603; N. D. Sen Gupta, Nucl.
Phys.  {\bf B4} (1967) 147; R. Wilson, ibid. B{\bf 68} (1974) 157;
Z. Tokuoka, Prog. Theor.  Phys. {\bf 37} (1967) 603; V. Fushchich, Theor.
Math. Phys. {\bf 9} (1970) 91; M.  T.  Simon, Lett.  Nuovo Cim.  {\bf 2}
(1971) 616; A. Raspini, Fizika {\bf B5} (1996) 159.

\bibitem{Shumovsky} A. S. Shumovsky y \"O. E. M\"ustecaplio\u{g}lu,
Phys. Rev. Lett. {\bf 80} (1998) 1202. Este trabajo presenta
la generalizaci\'on de los parametros de Stokes tomando en cuenta un modo
longitudinal de la radiaci\'on dipolar en la zona lejana.

}

\end{references}
\end{document}